# REDIS-BASED STREAMING ARCHITECTURE FOR ACCELERATOR BEAM INSTRUMENTATION DAQ SYSTEMS*

S. Joshi†, D. Steinkamp
Fermilab, Accelerator Division, Beam Instrumentation Department, Batavia, IL, USA

*Abstract*

The Fermilab Acceleraor Division, Beam Instrumentation Department, is always adopting modern and current software methodologies for complex DAQ architectures. This paper highlights the Redis Adapter (RA) as the key software component enabling high-performance, modular communication between digitizers and distributed control systems by leveraging Redis and containerization. The RA provides a unified, efficient interface between Redis-based data streams and consumer systems. In the legacy architecture, digitized data flowed through the custom, UDP-based Distributed Data Communication Protocol in the middle layer. In the current system, DDCP remains the ingestion path, while the RA serves as the decoupling layer. The proposed system replaces old VME digitizers with a SOM-based digitizer that communicates with Redis using the RA. The RA acts as both a performance-critical bridge and a protocol-agnostic adapter, ensuring compatibility with legacy control frameworks while enabling future scalability and modularity. This restructuring of the middle layer also helps the system achieve high throughput, reduce latency, and simplify the data path. Finally, we will demonstrate how RA is utilized in our two core products to deliver both legacy compatibility and future flexibility.

## INTRODUCTION

The Fermilab Beam Instrumentation (BI) Department monitors lifecycle diagnostics from end to end of beam systems across four major categories: Beam Current Monitors (BCM), Beam Loss Monitors (BLM), Beam Position Monitors (BPM), and Beam Profile Monitors (BProM). These systems provide critical data for accelerator stability, experimental optimization, and fault detection using instrumentation software that has evolved over decades to interface with VME-based digitizers and custom UDP protocols dating to the 1980s. While this traditional software lineage provides proven operational reliability, it increasingly limits the flexibility necessary for contemporary accelerator programs due to its tightly coupled nature. With the VME system's monolithic architecture and hardware constraints, there are development bottlenecks that make it no longer possible to meet the modern operational needs. The Proton Improvement Plan II (PIP-II) serves an urgency to modernization. The PIP-II will deliver world-record beam intensity for the DUNE neutrino experiment as the first U.S. particle accelerator to be built with substantial contributions from around the world and will facilitate diverse on-site physics programs over fifty years of operation. The extremely high-performance requirements of a superconducting linear accelerator, including never-before-seen intensities for beams, operational flexibility, and international collaboration workflows, reveal the fundamental limitations of legacy DAQ architectures. This paper discusses a strategic software solution for PIP-II that preserves existing control system expertise like ACNET (Accelerator Control Network) while enabling more options like EPICS (Experimental Physics and Industrial Control System) and architecture flexibility for the DAQ system. Through General Redis ACNET Front End (GRAFE) and General Redis EPICS Frontend (GREFE) implementations, we illustrate how RA bridges legacy systems with modern streaming architectures, providing evolutionary migration that simultaneously fulfils PIP-II's bold imperatives.

*Legacy Daq Architecture & Limitations*

Traditional VME-based systems feature tightly coupled components, including controller cards, digitizers, and PMC-UCD modules. Data from digitizers flows to control systems through specific VME drivers, kernels, and DDCP (Distributed Data Communication Protocol) implementations. Each consumer embeds device-specific protocol logic, duplicating effort and complicating deployments.

This architecture presents several major limitations:

- Scalability Constraints: Addition of new consumers or increasing sample counts required intrusive changes across multiple system layers and usually required co-ordinated downtime.
- Protocol Complexity: UDP-based implementation of DDCP was performant but required maintenance overhead and limited interoperability with modern containerized environments.
- Metadata Handling: Waveform metadata such as timestamps and sampling increments used ad hoc conventions, resulting in inconsistent data integration.
- Resource Allocation: Aging VME platforms required substantial sustaining engineering resources and diverted resources from new instrument development.
- Monolithic Coupling: Direct hardware-to-consumer connections led to brittle dependencies, making it harder to test, debug, or evolve.

Strict coupling between the data acquisition hardware and control system consumers in the legacy architecture imposed a maintenance burden that ultimately limited operational flexibility and development agility.

*Modern System Overview & Container Architecture*

The modernized architecture introduces clear separation of concerns through a Redis-based streaming system combined with comprehensive containerization and automated deployment pipelines. This integrated approach addresses both the technical limitations of legacy systems and the operational requirements for rapid development and reliable deployment.

## System Architecture Evolution

The decoupled producer-consumer framework: Modern architecture separates producers from consumers through Redis as a middle layer of the streaming [1]. Backend digitizers publish standardized data streams, and front-end systems join relevant channels in isolation to ensure system composition and evolution.

Protocol Adaptive Model: Redis Adapter is a universal translation layer that supports different protocols at the same time. The legacy DDCP is integrated along similar lines with the modern Redis-native components, while both ACNET and EPICS consumers use the same data streams through their own protocol-specific adapters.

Scalable Data Distribution: Redis Streams allows for high-performance, persistent data distribution and goes from single-consumer to large multi-consumer configurations, as well as its high-performance consumption patterns. The streaming model itself inherently allows real-time control applications, archived data, analytics pipelines, monitoring, and analytics infrastructure management to run in sync (rather than for consumers to coordinate). Operational flexibility: with containerized deployment a system can be quickly reconfigured, components are scaled independently, and maintenance is easy. New diagnostic features can be added without causing any new issues in the current systems, while some old ones can get started with a legacy interface.

## Data Flow Architecture

Figure 1 illustrates the fundamental data flow paradigm that underlies the modern architecture.

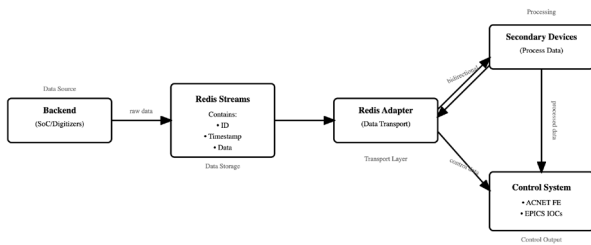

Figure 1: Overview of DAQ data flow.

The Redis Adapter serves as the central transport layer in this pipeline:

- Data ingestion: Backend systems (SoC/Digitizers) create raw data streams, which are streamed to Redis Streams that include structured data with fields like ID, timestamp, and payload, thus eliminating the full 48-byte header overhead of DDCP.
- Transport Layer: The Redis Adapter functions as a high-performance data transport service that consumes timestamped data from Redis streams and then performs protocol translation with no version compatibility issues such as those in DDCP evolution.
- Dual Output Paths:
- Direct Control Path: Streams data directly to Control Systems (ACNET FE, EPICS IOCs) for real-time accelerator control
- Processing Path: Processed data is routed through secondary devices and then analyzed or logged before final control output.

## Target System Architecture

Figure 2 provides a full target architecture with three tiers (that updates the full DAQ system), while maintaining compatibility with existing control frameworks.

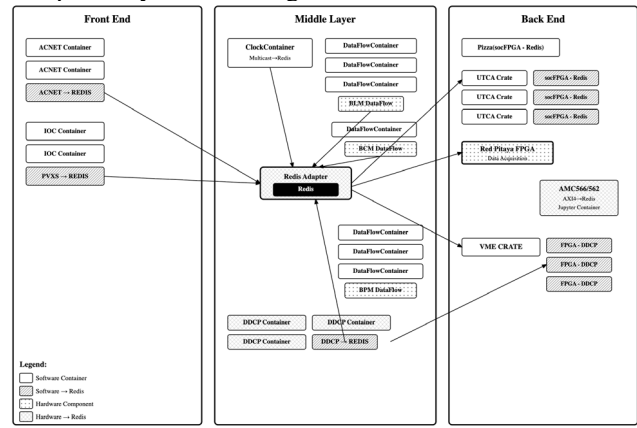

Figure 2: System Architecture Diagram.

Front End Tier: ACNET Containers and IOC Containers interface with the Redis Adapter via specialized protocol interfaces, thus bypassing DDCP's monolithic client-server dependencies.

Middle Layer: The Redis Adapter acts as the centralized hub with specialized containers, including the Clock Container for timing synchronization, Dataflow Containers for high-throughput processing, and DDCP Containers for legacy protocol support.

Back End Tier: The SOM-based solutions (Pizza box FPGA, socFPGA, and Red Pitaya) interface directly with Redis, while legacy VME systems connect through adapter containers.

## Container Infrastructure & DevOps Pipeline

Harbor Registry: The instrumentation project uses Harbor as container registry for centralized image management, organizing components like ACNET front-end, backend, and any secondaries under the specific namespace with vulnerability scanning and access control.

GitHub Actions CI/CD: Automated workflows trigger on main branch pushes and pull requests, running on Fermilab infrastructure (server) with secure Harbor authentication [2]. Multi-component builds use dynamic tagging, combining branch names, run numbers, and commit hash for precise version tracking.

Production Pipeline: The system builds and pushes both the latest and versioned tags to Harbor, enabling development workflows and production deployments while maintaining security through GitHub secrets management and SSH authentication for private repositories.

This containerized infrastructure significantly reduces operational complexity compared to older static configuration management while enabling the modern development needed for PIP-II and future projects' requirements.

Figure 3 demonstrates scalable deployment supporting multiple projects with shared infrastructure services, including Docker networks, container registries, and orchestration systems.

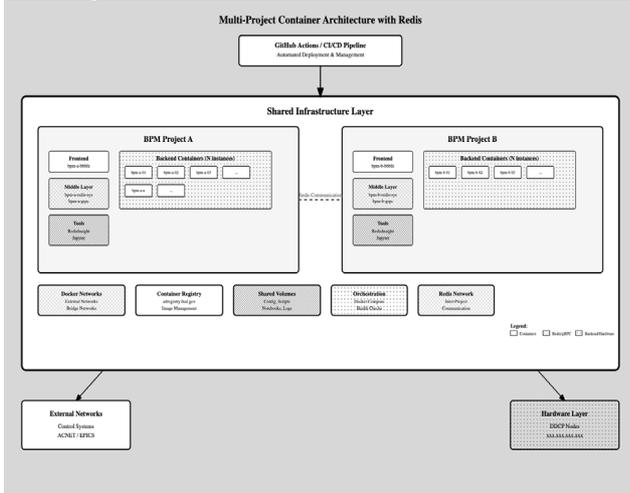

Figure 3: Shows multiple project container architecture.

## REDIS ADAPTER DESIGN AND IMPLEMENTATION

RA is a middle layer that integrates various system components and Redis database operations to process data in an organized way. The adapter, which is built on top of the C++ Redis client redis++, takes advantage of optimized internals so that data operations can be performed efficiently, while supporting advanced Redis attributes such as pipelining, transactions, and asynchronous commands.

Transport & Data Model: All backend→RA communication is done using Redis Streams for their native ordering, persistence, and consumer group feature [3]. The stream elements are equipped with a nanosecond-resolution timestamp using the Redis element ID in the standard millisecond-nanosecond structure (ms-ns). It allows for microsecond-precision timing without extra payload overhead. Writers control stream depth dynamically, single-element streams for scalar values optimize for low latency, and small sliding windows for high-frequency waveform data balance memory use and data availability. Binary payloads have a little-endian format (integers, IEEE-754 floats/doubles) to allow cross-platform application, and auxiliary fields contain waveform metadata.

Implementation to Customers & Reliability: RA is developed C++ library with redis++ client library based on its performance and strong connection control [4]. This implementation uses Redis pipelining for bulk operations, blocking reads to handle low-latency events, and auto-connection repair for high availability.

Authentication & Security: Enabling support of Redis AUTH and TLS encryption ensures production deployments are safe.

Namespacing & Key Management: The data organization is based on a hierarchical {BaseKey}: Subkey schema, which directly corresponds to the machine, the system, its physical instrument structure, the device, and its properties. The below Fig. 4 shows how the Booster BPM project will set up its Redis keys structure. RA makes it possible to maintain mapping tables between Redis keys and consumer identifiers (ACNET device properties, EPICS process variables), preserving both timestamp accuracy and metadata during the translation step.

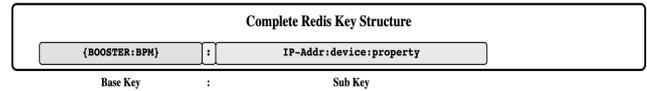

Figure 4: Redis key management.

Translation & Control Path: The RA translation engine can translate binary payloads to consumer-typed values and maintains data fidelity and timing information. The control path has two directions: monitoring and controlling.

Reading Path: The data stream from Redis with configurable update rates and data windowing is to consumers.

Write Path: SETs from consumers write to set up Redis PutKey streams that give access to the service by the same architecture.

Confirmation: Optional confirmation should be done to a designated PutKey before acceptance for control operation end, allowing reliable closed-loop operation for mission-critical control functions.

Figure 1 shows the flow of data between the backend application systems consuming Redis streams through the Redis Adapter and receiving the inputs for real-time control and the processing of the data to develop a flexible data analysis workflow.

## GRAFE (ACNET INTEGRATION)

GRAFE (General Redis Acnet Front Ends) provides a comprehensive ACNET integration path through containerized deployment and automated code generation, enabling rapid development of Redis-based ACNET front ends as shown in Fig. 5.

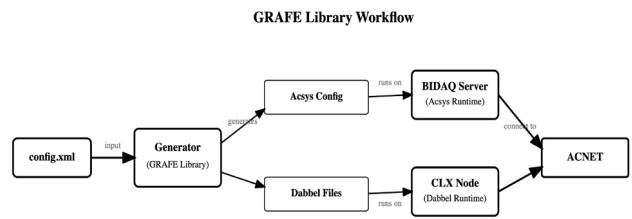

Figure 5: Demonstration on how this library works and preset data over ACNET.

*Configuration-Driven Development*

XML Configuration Workflow: GRAFE uses a declarative Config.xml file defining device, process, and parameter specifications with Redis key mappings. The XMLGenerator web interface provides form-based configuration generation with XML parsing capabilities for iterative development.

Hierarchical Configuration Structure: The configuration schema supports device-level settings (machine, maintainer, node), process-level parameters (processID, baseKey, serverAddress), and parameter-specific definitions, including Redis read/write keys, data types, array dimensions, and ACNET transform coefficients.

Code Generation Pipeline: The integrated generator validates XML configurations and produces Generated.dabbel file for ACNET parameter definitions and acsysconfig for runtime configuration, ensuring consistency between Redis data structures and ACNET infrastructure.

*Container-Based Deployment*

Adregistry Integration: GRAFE deploys through Docker containers with automated image management via container registry [5]. The workflow includes container building, image tagging, registry push operations, and deployment through docker-compose configurations.

Network Integration: Production deployment uses Docker networking with static IP assignment and hostname management for ACNET registration. The deployment supports both development environments with internet access and production control network isolation.

Automated Build Process: The container build process integrates code generation, compilation, and configuration file creation within the containerized environment, enabling consistent deployment across different operational environments.

*Parameter Management & Features*

ACNET Parameter Support: Each process supports up to 256 parameters with comprehensive ACNET feature integration, including transform coefficients (multiplier, divisor, offset), frequency constraints (ftpFrequencyMax, snpFrequencyMax), caching capabilities, and save list management.

Redis Integration: Parameters map to Redis streams through configurable read and write keys, supporting both scalar and array data types with optional timestamp positioning (leading) for waveform applications.

Operational Features: The system supports parameter logging at device, process, or parameter levels; automatic parameter distribution across multiple threads for scalability; and runtime frontend restart capabilities for operational flexibility.

This automated approach reduces ACNET front-end development cycles from weeks to hours while maintaining full compatibility with existing accelerator control infrastructure and operational procedures.

## GREFE (EPICS INTEGRATION)

GREFE (General Redis EPICS Front End) provides parallel EPICS integration capabilities using modern PVXS libraries for high-performance process variable operations. The path of how this library works is shown in Fig. 6. The EPICS integration approach is currently under active development and continues to evolve as new operational requirements emerge.

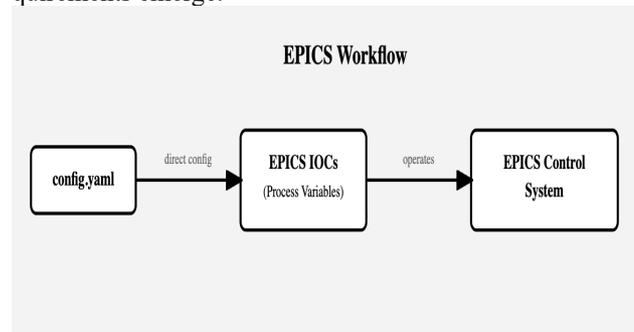

Figure 6: Shows how the EPICS front-end library works and flows to the control system.

*PVXS-Based Architecture*

Modern EPICS Implementation: GREFE utilizes the PVXS server framework, departing from traditional EPICS IOC architecture by eliminating database (DBD) and record (DB) files in favor of direct YAML configuration. Process variables are specified in a single config.yaml file rather than through traditional database scanning mechanisms.

Redis-Driven Updates: Instead of conventional database scanning, Redis explicitly updates process variables via notification when data changes. This approach provides more responsive and efficient data updates compared to periodic scanning while maintaining EPICS client compatibility.

Client Interface Compatibility: GREFE maintains standard EPICS behavior where client put operations invoke onPut handlers like traditional record processing, while get operations occur transparently without custom handling requirements.

*YAML Configuration Management*

Hierarchical Configuration: The config.yaml structure includes IOC-level settings (IOCName, PVBase, Redis Base, Redis connection parameters) and PV-specific definitions within a PVList array. Each process variable specifies name, description, Redis key mappings, data types, and array dimensions.

Redis Key Integration: Process Variables map to Redis streams through configurable GetKey and PutKey specifications. The system supports read-only PVs (GetKey only), write-only PVs (PutKey only), and bidirectional PVs with optional setting confirmation when keys differ.

Data Type Support: GREFE supports comprehensive EPICS data types including Boolean, string, signed/unsigned integers (8/16/32/64-bit), and floating-point values (32/64-bit) with both scalar and array configurations.

This run-time configuration driven approach demonstrates the Redis Adapter's flexibility in supporting diverse control system integration patterns while enabling rapid prototyping and iterative development of EPICS interfaces as operational requirements continue to evolve.

# FUTURE WORK & CONCLUSION

*Planned Enhancements*

GREFE Evolution: Unlike the mature GRAFE implementation, GREFE represents an evolving approach to EPICS integration that adapts to emerging operational needs and requirements. The system design prioritizes flexibility to accommodate future EPICS ecosystem developments and changing accelerator control requirements.

Python Integration: Development of a Python wrapper for the Redis Adapter will allow us to leverage the rich Python ecosystem, enabling rapid development of analytics and machine learning integrations. Python's simplicity will facilitate integration of new features while maintaining interoperability with the existing C++ foundation, supporting diverse use cases from data analytics to third-party system integration.

Horizontal Scaling & Integration: Usage of any new platform of streaming frameworks will enable real-time data analysis and machine learning much more easily.

# CONCLUSION

Redis Adapter architecture solves the scalability and maintenance issues of legacy accelerator DAQ systems. It decouples data producers and consumers by standardizing on Redis Streams while ensuring the essential performance for particle accelerator operations. The dual-path solution of GRAFE and GREFE is compatible with existing control systems and modern streaming architectures. GRAFE's robust ACNET integration and GREFE's evolving design demonstrate flexibility at multiple control levels. The real-time performance data on deployments proves the architecture meets the strictest operational needs. Containerized deployment makes deployment simpler and enhances observability and debugging. The Redis Adapter lays a solid foundation for future software architectures, which will enable Fermilab to update its accelerator infrastructure for PIP-II and future accelerator programs.

# ACKNOWLEDGMENTS

We thank our dedicated BI software development team whose collaborative efforts made the Redis Adapter architecture possible. The successful implementation spanning the core Redis Adapter library, GRAFE and GREFE integration paths, and containerized deployment infrastructure represents our collective expertise and shared commitment to modernizing accelerator instrumentation software. We acknowledge the Beam Instrumentation and Controls teams at Fermilab for their technical contributions and operational support.

# REFERENCES

[1] R. Santucci, J. Diamond, N. Eddy, A. Semenov, D. Voy, "A Modern Ethernet Data Acquisition Architecture for Fermilab Beam Instrumentation," Fermilab, Batavia, IL, 2022, Report No. FERMILAB-POSTER-22-215-AD [6].

[2] "GitHub Actions Documentation," https://docs.github.com/en/actions

[3] "Redis Documentation," https://redis.io/docs/

[4] Fermi Accelerator Division, "Redis Adapter: C++ adapter which wraps redis++ to communicate to the instrumentation Redis database," GitHub, 2024, https://github.com/fermi-ad/redis-adapter

[5] "Docker Documentation," https://docs.docker.com/reference/dockerfile/

[6] "Preparation of Papers for JACoW Conferences," http://jacow.org/

\*Work supported by Fermi Forward Discovery Group, LLC under United States Department of Energy.
†sjoshi@fnal.gov